\documentclass[3p, times]{elsarticle}

\usepackage{amssymb}

 \biboptions{compress}

\usepackage[figuresright]{rotating}

\begin{document}

\begin{frontmatter}



\title{Arithmetic Brownian motion subordinated by tempered stable and inverse tempered stable processes}


\author{ Agnieszka Wy{\l}oma{\'n}ska}

\address{Hugo Steinhaus Center, Institute of Mathematics and Computer Science,\\     
    Wroclaw University of Technology, Poland\\
    agnieszka.wylomanska@pwr.wroc.pl}

\begin{abstract}
In the last decade the subordinated processes have become popular and found many practical applications. Therefore in this paper we examine two processes related to time-changed (subordinated) classical Brownian motion with drift (called arithmetic Brownian motion). The first one, so called normal tempered stable, is related to the tempered stable subordinator, while the second one - to the inverse tempered stable process. We compare the main  properties (such as probability density functions, Laplace transforms, ensemble averaged mean squared displacements) of such two subordinated processes and propose the parameters' estimation procedures. Moreover we calibrate the analyzed systems to real data related to indoor air quality.  
 
\end{abstract}
\end{frontmatter}
\section{Introduction}
Processes based on the Brownian motion were considered in many aspects and have found various practical applications \cite{bach,bm1,bm2,bm3,bm4,bm5,bm6,bm7}. But the assumption of normality for the observations seems not to be reasonable in the number of examined  phenomenon. Therefore in many Gaussian models the Brownian motion is replaced by its various modifications. One of the simples modification is the extension of Gaussian  by another distribution, for example  the L\'evy-stable one. Processes based on the stable distribution are very useful in modeling data that exhibit fat tails. For example, the  classical Ornstein–-Uhlenbeck process was extended to the stable case and analyzed in \cite{ou1,ou2} as a suitable model for  financial data description, see also \cite{mojasia}. The another possibility of modification for Brownian-type processes is introduction a time-changed Brownian models. This extension is related to replacement of real time in Brownian systems by non-decreasing  L\'evy process (called subordinator), that in this case plays a role of random (operational) time. The new process is called subordinate. The idea of subordination was introduced in 1949 by Bochner \cite{boch} and expounded in his book in \cite{boch2}. The theory of subordinated processes is also explored in details in \cite{sato}.

The subordinated processes based on the diffusive Brownian motion were considered in many disciplines. The Brownian motion subordinated by gamma process, so called variance gamma, is analyzed in \cite{vg1} in the context of option price modeling. The other applications and  main characteristics of such system one can find in \cite{vg2}. Subordination of Brownian motion by the inverse Gaussian process is called the normal inverse Gaussian (NIG)  and was considered for example in \cite{bnnig} and proposed to modeling turbulence and financial data. Applications of NIG processes to asset returns are also shown in \cite{rafal1} while analysis of real environmental data by using NIG distribution is presented in \cite{moja_smoluch}. Let us mention that  the Brownian motion driven by  strictly increasing L\'evy subordinator is also a L\'evy process, moreover when the subordinator is temporally homogeneous Markov  then the subordinate has also this property, \cite{sato}. 

Another possibility of subordination is replacement the real time in Brownian diffusion by  inverse subordinators and processes that arise after this transformation are called anomalous diffusion. In the domain of anomalous diffusion the typical approach is based on continuous time random
walk (CTRW), \cite{a23,a30}, and subordinated L\'evy processes can be treat as a limit in distribution of CTRW,
\cite{magwerwer}. The key issue in the framework of CTRW as well as in subordination technique
is the waiting-times distribution corresponding to observed  constant time periods \cite{orzelwyl}.  In the last decade the anomalous diffusion processes were analyzed by various number of authors in many disciplines. For example the subordinated Brownian motion driven by inverse L\'evy-stable subordinator was considered in \cite{mojasia,magwerwer,magsam,stanis2}, the inverse tempered stable subordinator was examined in \cite{orzelwyl,stanis,maggaj1,maggaj2} while inverse gamma process was mentioned in \cite{janczurawylomanska2011}. The general case of L\'evy processes that can play a role of inverse subordinators were explored for example in \cite{sbm1,magsam1,woy}.  

In this paper we examine two processes related to subordinated Brownian motion. The first one, so called normal tempered stable \cite{ole,temp3}, is a Brownian motion with drift (called arithmetic Brownian motion, ABM) driven by tempered stable subordinator, while the second one is an ABM subordinated by inverse tempered stable subordinator. The  tempered stable  processes are  extension of the  $\alpha-$stable L\'evy systems but possess also  the properties of Gaussian models, therefore in the last few years they have become popular and very useful in description of many real data, \cite{ou2,temp3,temp1}.  We compare the main statistical properties of the ABM driven by tempered stable and inverse tempered stable subordinator.  Moreover in two considered cases we propose the parameters' estimation procedures and validate them. In order to illustrate the theoretical results we calibrate the examined processes to real data related to indoor air quality. 

The rest of the paper is organized as follows: In section \ref{TSS} we introduce the ABM and tempered stable subordinator. We present the main properties of those processes and define the time-changed ABM driven by  tempered stable process. For this system we also examine the main statistical characteristics, such as probability density function, Laplace transform and ensemble averaged mean squared displacement that can be an useful tool to distinction between diffusion and anomalous diffusion models. In this section we propose also the estimation procedure based on the distance between theoretical and empirical Laplace transforms and validate it. In section \ref{ITSS} we examine inverse tempered stable subordinator and its main statistical properties as well as define the  ABM driven by inverse tempered stable process. In this section we also present the estimation procedure for unknown parameters. In section \ref{appl} we model  real data sets  related to indoor air quality by using the mentioned subordinated processes. Last section contains conclusions.

\section{Arithmetic Brownian motion with tempered stable subordinator}\label{TSS}
\subsection{Arithmetic Brownian motion}
The arithmetic Brownian motion (ABM)  is a process $\{X(t),~t\geq 0\}$ defined by \cite{orzelwyl,magorzelweron}:
\begin{eqnarray}\label{abm}
dX(t)=\beta dt +dB(t),
\end{eqnarray}
where $\{B(t),~t\geq 0\}$ is a classical Brownian motion and $\beta>0$. The solution of equation (\ref{abm}) takes the form
\begin{eqnarray*}
X(t)=X(0)+\beta t+ B(t).
\end{eqnarray*}
In the further analysis we assume $X(0)=0$ with probability one and in this case the process defined above has Gaussian distribution with  mean $\beta t$ and  variance $t$. It is called also Brownian motion with drift, \cite{orzelwyl} and first of all was  used to description of stock prices \cite{bach,pit}.  It is an extension of the classical Brownian motion therefore its modifications have found many other practical applications, like diffusion in liquids modeling, \cite{liq} and description of hydrology time series \cite{mojakrzysiek}.
\subsection{Tempered stable subordinator}
The tempered stable subordinator $\{T(t),~t\geq0\}$ is  a strictly increasing L\'evy process with tempered stable increments, i.e. with the following Laplace transform, \cite{maggaj1, maggaj2}:
\begin{eqnarray}\label{tss}
<e^{-zT(t)}>=e^{t(\lambda^\alpha-(\lambda+z)^{\alpha})},~\lambda>0, 0<\alpha<1.
\end{eqnarray}
When $\lambda=0$, then $\{T(t)\}$ becomes totally skewed $\alpha-$stable L\'evy process. The probability density function (pdf) of tempered stable subordinator can be express in the following form:
\begin{eqnarray*}
f_{T(t)}(x)=e^{-\lambda x+\lambda^{\alpha}t}f_{U(t)}(x),
\end{eqnarray*}
where $\{U(t),~t\geq0\}$ is a totally skewed  $\alpha-$stable L\'evy motion with the stability index $\alpha$, \cite{orzelwyl,baumer}. 
By using tail approximation of stable density, \cite{nolan}, we obtain the following:
\begin{eqnarray*}f_{T(t)}(x)\sim 2\alpha c_{\alpha}e^{-\lambda x+\lambda^{\alpha}t} t^{\alpha}x^{-(\alpha+1)},~x\rightarrow\infty\end{eqnarray*}
for some constant $c_{\alpha}$. Therefore the right tail  can be approximated by:
\begin{eqnarray*}
1-F_{T(t)}(x)\sim e^{-\lambda x+\lambda^{\alpha}t} \left(\frac{t}{x}\right)^{\alpha},~~x\rightarrow\infty,
\end{eqnarray*}
where $F_{T(t)}(\cdot)$ is the distribution function of $T(t)$. Let us mention that for $\lambda=0$ (the $\alpha-$stable case), the above formula reduces to $1-F_{U(t)}(x)\sim  x^{-\alpha}$ for fixed $t$.
\subsection{ABM with tempered stable subordinator}
\begin{figure}[tb]
\begin{center}
\includegraphics[width=9cm]{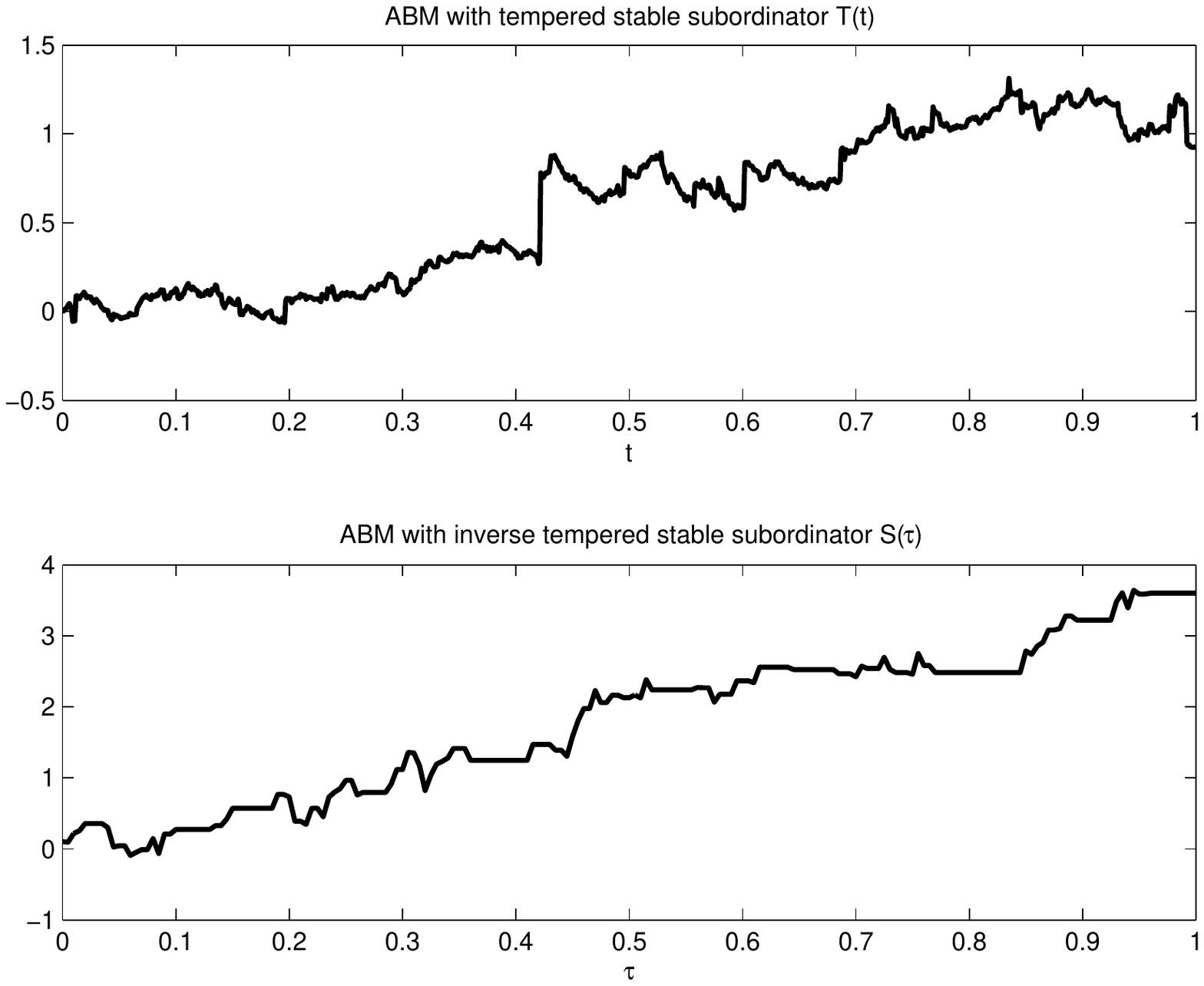} \caption{The sample trajectory of the process $\{Y_T(t),~t\geq0\}$ subordinated by tempered stable subordinator (top panel) and process $\{Y_S(t),~t\geq 0\}$ subordinated by inverse tempered stable subordinator (bottom panel) with parameters $\alpha=0.8$, $\lambda=1$ and $\beta=1$ on the interval $[0,1]$.}\label{fig1}
\end{center}
\end{figure}

The  ABM driven by tempered stable subordinator is the process $\{Y_T(t),~t\geq 0\}$ defined as follows:
\begin{eqnarray*}
Y_T(t)=X(T(t)),
\end{eqnarray*}
where $\{X(t)\}$ is ABM defined in (\ref{abm}) and $\{T(t)\}$ is strictly increasing L\'evy process with tempered stable increments with the Laplace transform given in (\ref{tss}). We assume those two processes are independent.\\
Using the form of the solution of equation (\ref{abm}) we can write the exact formula for the process $\{Y_T(t)\}$, namely:
\begin{eqnarray}\label{yt}
Y_T(t)=B(T(t))+\beta T(t).
\end{eqnarray}
The process $\{Y_T(t)\}$ is known in the literature as normal tempered stable, \cite{ole} and the algorithm for generating its sample
trajectories  in points $t_1,t_2,...,t_n$ proceeds follows:
\begin{enumerate}
\item Simulate the increments of subordinator $d T_i=T(t_i)-T(t_{i-i})$ with the initial value $T_0=0$. In order to do this use the simple algorithm presented for example in \cite{maggaj1}.
\item Simulate $n$ standard Gaussian random variables $N_1, N_2,...,N_n$
\item Define $d Y_T(i)=N_i\sqrt{d T_i}+\beta d T_i$ and put $Y_T(t_i)=\sum_{k=1}^i\Delta Y_i$.
\end{enumerate}
The sample trajectory of the process $\{Y_T(t)\}$ with parameters $\alpha=0.8$, $\lambda=1$ and $\beta=1$ is presented on the top panel of Fig.\ref{fig1}.
\\

\begin{figure}[tb]
\begin{center}
\includegraphics[width=9cm]{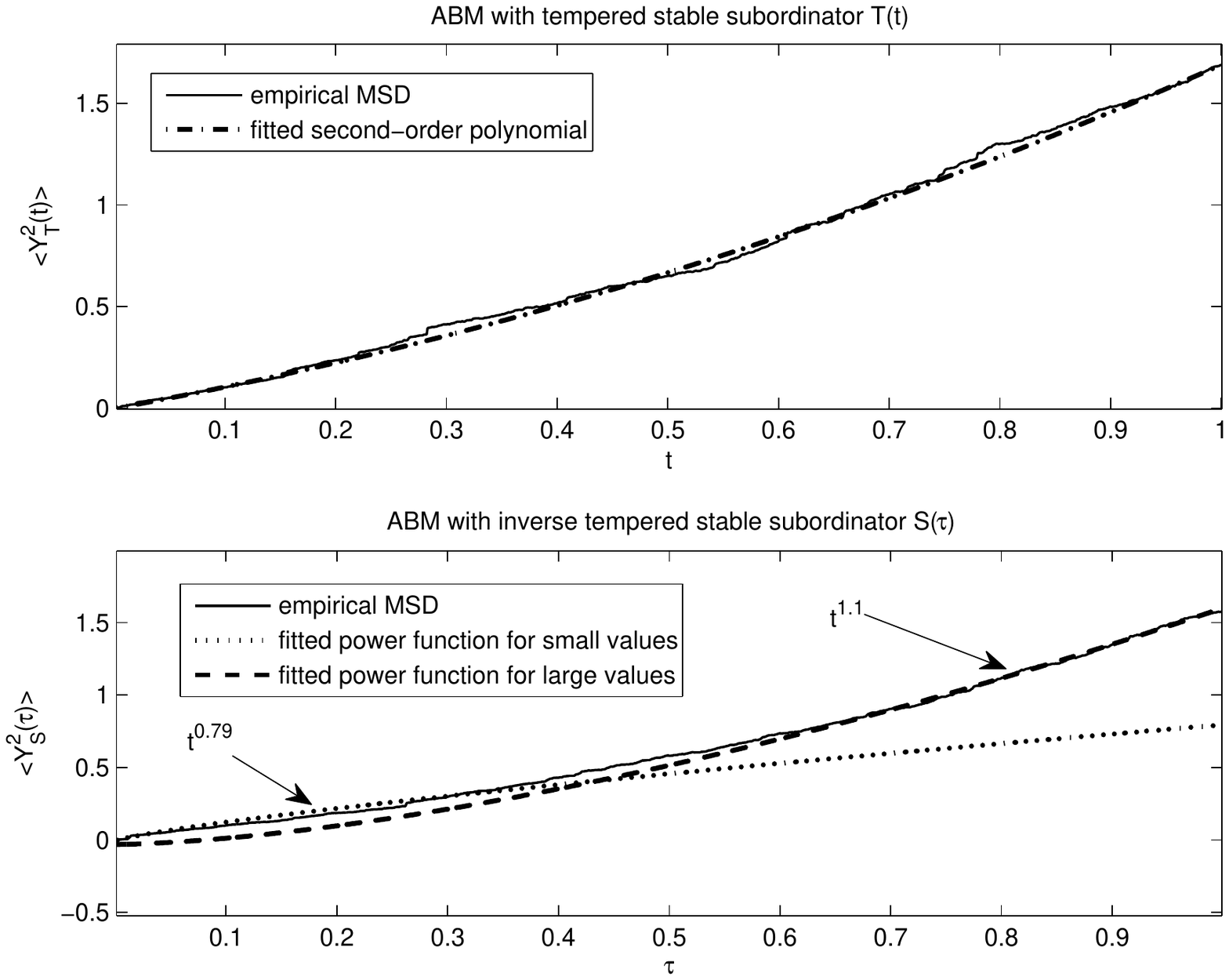} \caption{The ensemble averaged MSD  calculated on the basis of $1000$ trajectories  for the $\{Y_{T}(t)\}$ (top panel) and $\{Y_S(t)\}$ (bottom panel) processes with $\alpha=0.8$, $\lambda=1$ and $\beta=0.01$. On the top panel we present the fitted second-order polynomial while on the bottom panel - fitted power functions for small and large values.}\label{msd}
\end{center}
\end{figure}

On the basis of the formula for Laplace transform of general subordinated processes given in \cite{sato}  we can calculate this characteristic of the process $\{Y_T(t)\}$, namely:
\begin{eqnarray}\label{ltYT}<e^{-zY_T(t)}>=\exp\left\{t\left(\lambda^{\alpha}-\left(\lambda+\beta z-\frac{1}{2}z^2\right)^{\alpha}\right)\right\}.\end{eqnarray}
The pdf of the process $\{Y_T(t)\}$ defined in (\ref{yt}) can be calculated by using the following formula:
\begin{eqnarray*}
f_{Y_T(t)}(x)=\int_{0}^{\infty}f_{X(z)}(x)f_{T(t)}(z)dz,
\end{eqnarray*}
where $f_{X(z)}(\cdot)$ and $f_{T(t)}(\cdot)$ denote the pdfs of the ABM $\{X(t)\}$ and tempered stable subordinator $\{T(t)\}$, respectively. Therefore we obtain:
\begin{eqnarray*}
f_{Y_T(t)}(x)=\int_{0}^{\infty}\frac{1}{\sqrt{2\pi z}}e^{-(x-\beta z)^2/2z-\lambda z+\lambda^{\alpha}t}f_{U(t)}(z)dz,
\end{eqnarray*}
where $f_{U(t)}(\cdot)$ is the pdf of the totally skewed $\alpha-$stable L\'evy motion $\{U(t),~t\geq 0\}$ mentioned above. \\
On the basis of  formulas (\ref{yt}) and (\ref{ltYT}) we can also calculate the main characteristics such as mean and autocovariance function:
\[<Y_T(t)>=\beta t\alpha\lambda^{\alpha-1}\]\[cov(t,s)=<Y_T(t),Y_T(s)>-<Y_T(t)><Y_T(s)>=min(s,t)\left(\alpha\lambda^{\alpha-1}+\beta^2\alpha(1-\alpha)\lambda^{\alpha-2}\right).\]
We consider also one of the most popular characteristic of the process that is especially important  in real data analysis because it can be an useful tool for recognition between diffusion and anomalous diffusion models. This characteristic is called mean squared displacement (MSD). Here we consider  the ensemble averaged MSD that is defined as a second moment of a given process. Therefore for $\{Y_T(t)\}$ it takes the form, \cite{janczurawylomanska2011}:
\begin{eqnarray*}
<Y_T^2(t)>=\int_{0}^{\infty}x^2f_{Y_T(t)}(x)dx,
\end{eqnarray*}
where $f_{Y_T(t)}(\cdot)$ is a pdf of the process $\{Y_T(t)\}$. Let us emphasize, that the ensemble averaged MSD for real data can be calculated only in case when there  are available more than one trajectory. For the process $\{Y_T(t)\}$ defined in (\ref{yt}) the ensemble averaged MSD is a polynomial of the second order with respect to $t$:
\[<Y_T^2(t)>=t^2(\beta^2\alpha \lambda^{\alpha-1})^2+t\left(\alpha\lambda^{\alpha-1}+\beta^2\alpha(1-\alpha)\lambda^{\alpha-2}\right).\]
As we observe, in case of $\beta=0$, the ensemble averaged MSD  scales as $t$. In Fig. \ref{msd} (top panel) we present this characteristic  calculated on the basis of $1000$ trajectories for the ABM with tempered stable subordinator with $\alpha=0.8$, $\lambda=1$ and $\beta=0.01$. Moreover we also show the fitted second-order polynomial.
 
\subsection{Estimation procedure}
\begin{figure}[tb]
\begin{center}
\includegraphics[width=9cm]{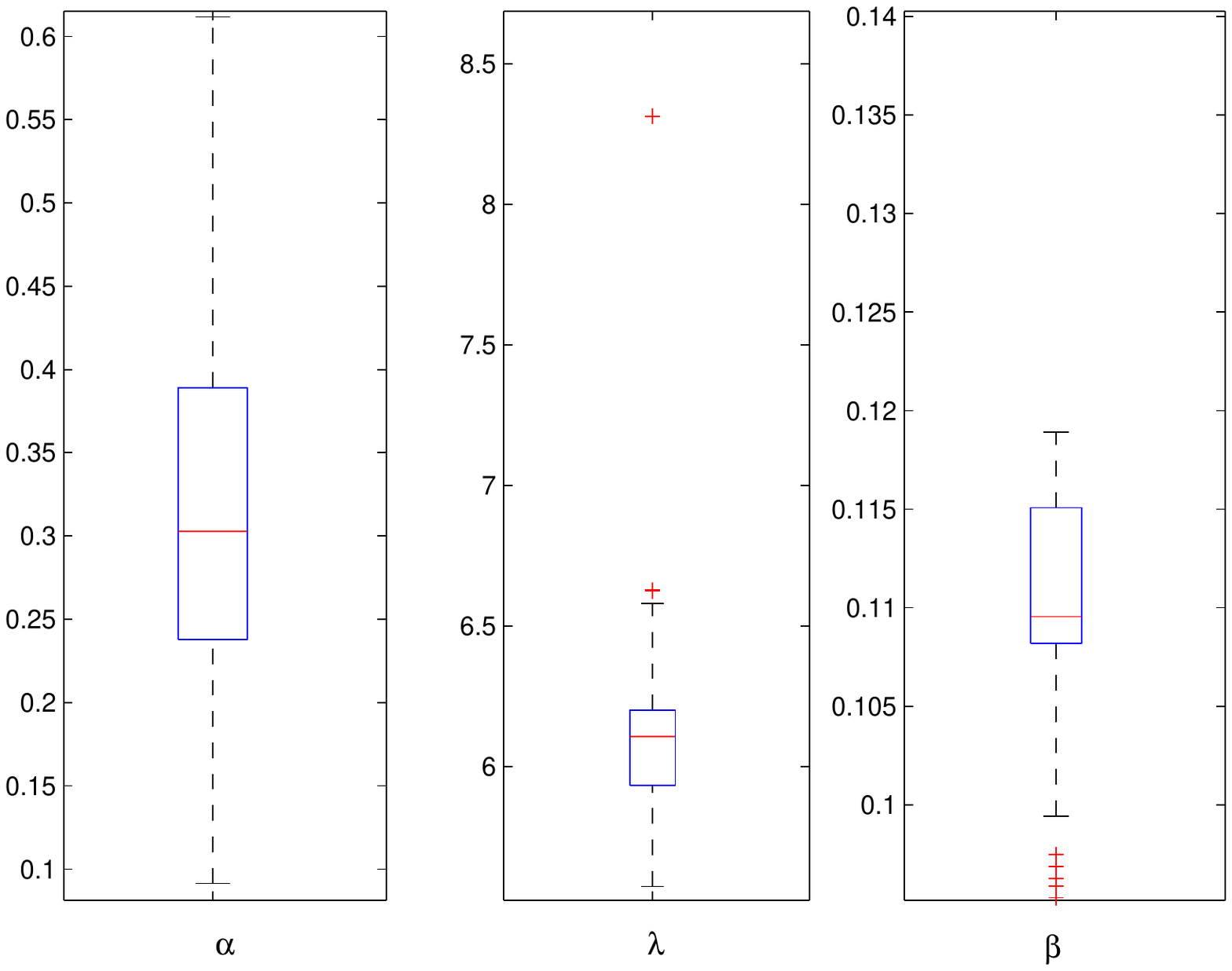} \caption{The boxplots of the values of estimators for parameters of ABM with tempered stable subordinator $\{Y_T(t)\}$. The values were calculated on the basis of $1000$ trajectories of length $1000$ each. The theoretical values are : $\alpha=0.26$, $\lambda=6$, $\beta=0.11$.}\label{boxplots}
\end{center}
\end{figure}

We propose to estimate  the parameters $\alpha$, $\lambda$ and $\beta$ by using the formula of the Laplace transform given in (\ref{ltYT}). Because  $\{Y_T(t)\}$ is a L\'evy process, \cite{sato}, therefore its increments are independent and have the same distribution with the following characteristic:
\[<e^{-z(Y_T(t+d)-Y_T(t))}>=\exp\left\{d\left(\lambda^{\alpha}-\left(\lambda+\beta z-\frac{1}{2}z^2\right)^{\alpha}\right)\right\}.\]
The estimation scheme for random sample $y_1,y_2,...,y_n$  proceeds as follows:
\begin{enumerate}
\item Calculate increments of the observations: $d y_i=y_{i+1}-y_i$ for $i=1,2,...,n-1$.
\item For the increments find the empirical Laplace transform by using the following formula:
\[\phi(z)=\frac{1}{n-1}\sum_{i=1}^{n-1}e^{-z\Delta y_i}.\]
\item By using the least squares method find the estimators $\hat{\alpha}$, $\hat{\lambda}$, $\hat{\beta}$ that satisfy:
\[(\hat{\alpha},\hat{\lambda},\hat{\beta})=min_{\alpha,\lambda,\beta}\left(\phi(z)-\exp\left\{\lambda^{\alpha}-\left(\lambda+\beta z-\frac{1}{2}z^2\right)^{\alpha}\right\}\right)^2.\]
\end{enumerate}
In order to show  the efficiency of the described method, in Fig. \ref{boxplots} we present the values of the estimated parameters for simulated  process $\{Y_T(t)\}$. To the analysis we take $1000$ trajectories of length $1000$. The theoretical values are: $\alpha=0.26$, $\lambda=6$, $\beta=0.11$. As we observe  the theoretical values are close to the estimated values (are between appropriate quantiles) that indicates the procedure works properly.

\section{ABM with inverse tempered stable subordinator}\label{ITSS}

In this section we examine the main characteristics of the ABM with driven by the inverse tempered stable subordinator. Such kind of systems were  considered in the literature in many aspects, \cite{orzelwyl,stanis,maggaj1,maggaj2,janczurawylomanska2011}, therefore  some  properties we present without proofs. 
 \subsection{Inverse tempered stable subordinator}
\begin{figure}[tb]
\begin{center}
\includegraphics[width=9cm]{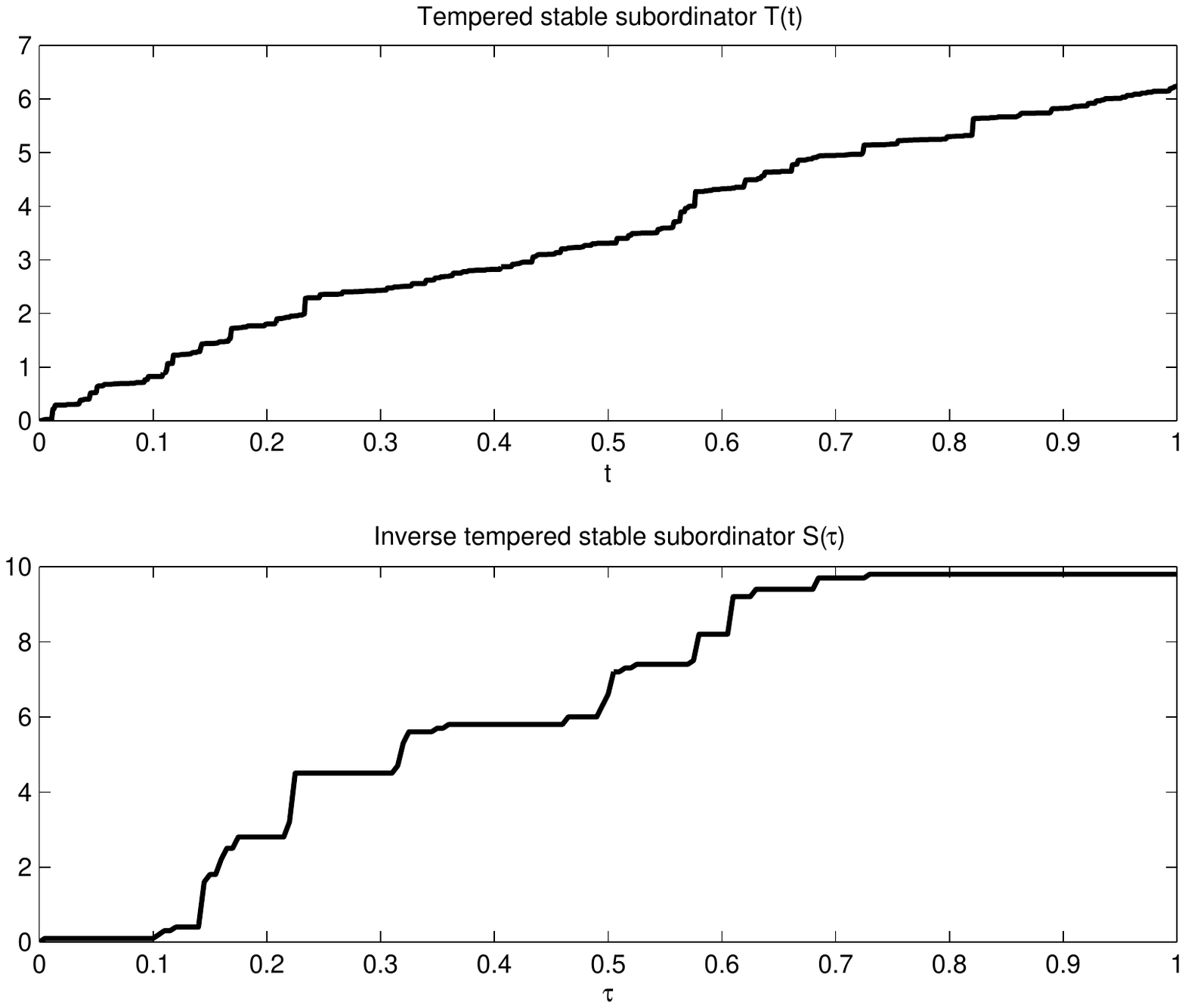} \caption{The sample trajectories of the tempered stable subordinator $\{T(t)\}$ (top panel) and inverse tempered stable subordinator $\{S(\tau)\}$ (bottom panel) for $\alpha=0.26$ and $\lambda=6$.}\label{porow1}
\end{center}
\end{figure}

The inverse tempered stable subordinator is the process $\{S(\tau),~\tau\geq 0\}$  defined as follows:
\begin{eqnarray*}
S(\tau)=\inf\{t>0:T(t)>\tau\},
\end{eqnarray*}
where $\{T(t),~t\geq 0\}$ is the tempered stable L\'evy process defined via its Laplace transform in (\ref{tss}). In Fig.\ref{porow1} we present the trajectories of the tempered stable subordinator $\{T(t)\}$ and corresponding tempered stable inverse subordinator  $\{S(\tau)\}$ for parameters $\alpha=0.26$ and $\lambda=6$. As we observe, the processes are non-decreasing but exhibit completely different behavior. 

Using the relation $P(S(\tau)\leq x)=P(T(x)>\tau)$ we can calculate the pdf of the process $\{S(\tau)\}$, namely:
\[f_{S(\tau)}(x)=-\frac{d}{d x}\int_{0}^{\tau}f_{T(x)}(u)du.\]
Therefore we get:
\begin{eqnarray*}f_{S(\tau)}(x)=-\frac{d}{d x}\int_{0}^{\tau}e^{-\lambda u+\lambda^{\alpha}x}f_{U(x)}(u)du,\end{eqnarray*}
where $\{U(t)\}$ is a totally skewed $\alpha-$stable L\'evy process  described in the previous section.\\
Because $f_{U(x)}(u)=\frac{1}{x^{1/\alpha}}f_{U(1)}(u/x^{1/\alpha})$, \cite{magdziarz0}, therefore we get:

\begin{eqnarray*}
f_{S(\tau)}(x)&=&-\frac{d}{d x}\int_{0}^{\tau}e^{-\lambda u+\lambda^{\alpha}x}\frac{1}{x^{1/\alpha}}f_{U(1)}(u/x^{1/\alpha})du\\
&=&-\int_{0}^{\tau}e^{-\lambda u+\lambda^{\alpha}x}\left[\lambda^{\alpha}\frac{1}{x^{1/\alpha}}f_{U(1)}(u/x^{1/\alpha})-\frac{1}{\alpha x^{1/\alpha+1}}f_{U(1)}(u/x^{1/\alpha})-\frac{u}{\alpha x^{2/\alpha+1}}f^{'}_{U(1)}(u/x^{1/\alpha})\right]du\\
&=&\left(\frac{1}{\alpha x}-\lambda^{\alpha}\right)\int_{0}^{\tau}e^{-\lambda u+\lambda^{\alpha}x}\frac{1}{x^{1/\alpha}}f_{U(1)}(u/x^{1/\alpha})du+\int_{0}^{\tau}e^{-\lambda u+\lambda^{\alpha}x}\frac{u}{\alpha x^{2/\alpha+1}}f^{'}_{U(1)}(u/x^{1/\alpha})du\\
&=&\left(\frac{1}{\alpha x}-\lambda^{\alpha}\right)\int_{0}^{\tau/x^{1/\alpha}}e^{-\lambda ux^{1/\alpha}+\lambda^{\alpha}x}f_{U(1)}(u)du+\frac{1}{\alpha x}\int_{0}^{\tau/x^{1/\alpha}}e^{-\lambda u x^{1/\alpha}+\lambda^{\alpha}x}uf^{'}_{U(1)}(u)du\\
&=&\frac{1}{\alpha x}e^{\lambda^{\alpha} x}\int_{0}^{\tau/x^{1/\alpha}}e^{-\lambda u x^{1/\alpha}}(f_{U(1)}(u)+uf^{'}_{U(1)}(u))du-\lambda^{\alpha}\int_{0}^{\tau/x^{1/\alpha}}e^{-\lambda ux^{1/\alpha}+\lambda^{\alpha}x}f_{U(1)}(u)du\\
&=&\frac{1}{\alpha x}e^{\lambda^{\alpha} x}\int_{0}^{\tau/x^{1/\alpha}}(e^{-\lambda u x^{1/\alpha}}uf_{U(1)}(u))'+\lambda e^{-\lambda u x^{1/\alpha}}x^{1/\alpha}uf_{U(1)}(u)du-\lambda^{\alpha}\int_{0}^{\tau/x^{1/\alpha}}e^{-\lambda ux^{1/\alpha}+\lambda^{\alpha}x}f_{U(1)}(u)du.
\end{eqnarray*}

As a final result we obtain:
\begin{eqnarray}\label{gestoscS}
f_{S(\tau)}(x)=\frac{1}{\alpha x}\tau f_{T(x)}(\tau)+\lambda\frac{1}{\alpha x}\int_{0}^{\tau}uf_{T(x)}(u)du-\lambda^{\alpha}\int_{0}^{\tau}f_{T(x)}(u)du.
\end{eqnarray}

In case of $\lambda=0$ we have:
\[f_{S(\tau)}(x)=\frac{1}{\alpha x}\tau f_{U(x)}(\tau),\]
that coincides with the result presented in \cite{magdziarz0} for $\alpha-$stable case. For large $\tau$ the density given in (\ref{gestoscS}) tends to:
\[f_{S(\tau)}(x)\sim\frac{1}{\alpha x}\tau f_{T(x)}(\tau)+\lambda\frac{1}{\alpha x}<T(x)>-\lambda^{\alpha}=\frac{1}{\alpha x}\tau f_{T(x)}(\tau).\]
Similar, for small $\tau$ we get:
\[f_{S(\tau)}(x)\sim\frac{1}{\alpha x}\tau f_{T(x)}(\tau).\]
On the basis of equation (\ref{gestoscS}) we can calculate also the Laplace transform of $\{S(\tau)\}$, namely:
\begin{eqnarray*}
<e^{-zS(\tau)}>=\int_{0}^{\infty}e^{-zx}f_{S(\tau)}(x)dx.
\end{eqnarray*}
However such derivation requires numerical approximations.

 \subsection{ABM driven by inverse tempered stable subordinator}

The ABM driven by the inverse tempered stable subordinator is defined as follows:
\begin{eqnarray*}
Y_S(\tau)=X(S(\tau))=\beta S(\tau)+B(S(\tau)),
\end{eqnarray*}
where $\{X(t)\}$ is the classical ABM explored in section \ref{TSS}.  The simulation procedure of the process $\{Y_S(\tau)),~\tau\geq 0\}$ is completely described in \cite{maggaj1}. The sample trajectory of the process is presented in Fig.  \ref{fig1} (bottom panel). To the simulation we take the following values of the parameters:  $\alpha=0.8$, $\lambda=1$ and $\beta=1$. Let us remind that on the top panel we present the sample trajectory of the ABM with tempered stable subordinator $\{T(t)\}$. For  $\{Y_S(\tau)\}$ we observe  constant time periods that is typical for subdiffusive processes. This behavior is not visible for the $\{Y_T(t)\}$.\\
The Laplace transform of the ABM driven by inverse tempered stable subordinator $\{S(\tau)\}$ can be calculated by using the following formula:
\[<e^{-zY_S(\tau)}>= \int_{0}^{\infty}e^{-(\beta z-\frac{1}{2}z^2)x}f_{S(\tau)}(x)dx,\]
where $f_{S(\tau)}(\cdot)$ is given in (\ref{gestoscS}).
The pdf $f_{Y_S(\tau)}(x)$ of the process $\{Y_S(\tau)\}$ satisfies the generalized Fokker-Planck equation:
\[\frac{d f_{Y_S(\tau)}(x)}{d \tau}=\left[-\beta \frac{d}{d x}+\frac{1}{2}\frac{d^2}{d^2x}\right]\Phi (f_{Y_S(\tau)}(x)).\]
where $\Phi$ is an operator defined in \cite{orzelwyl}. On the other hand we can use the formula for pdf of subordinated processes given in \cite{sato}:
\[ f_{Y_S(\tau)}(x)=\int_{0}^{\infty}f_{X(z)}(x)f_{S(\tau)}(z)dz.\]
Therefore we obtain the following:
\[f_{Y_S(\tau)}(x)=\int_{0}^{\infty}\frac{1}{\sqrt{2\pi z}}e^{-(x-\beta z)^2/2z}f_{S(\tau)}(z)dz,\]
The  mean of the process $\{Y_S(\tau)\}$ is given by, \cite{janczurawylomanska2011}:
\[<Y_S(\tau)>=\beta <S(\tau)>=\beta\int_{0}^{\tau}e^{-\lambda u}u^{\alpha-1}E_{\alpha,\alpha}((\lambda u)^{\alpha})du,\]
where $E_{\alpha,\beta}(z)$ is a generalized Mittag-Leffler function defined as follows, \cite{mittag}:
\[E_{\alpha,\beta}(z)=\sum_{k=0}^{\infty}\frac{z^k}{\Gamma(\alpha k+\beta)}.\]
For small $\tau$, $<Y_S(\tau)>\sim \tau^{\alpha}$, while for large values of $\tau$ the mean tends to $\tau$, \cite{stanis}.
Moreover the ensemble averaged MSD takes the form, \cite{magdziarzinfinity}:
\[<Y_S^2(\tau)>=\beta^2<S^2(\tau)>+<B^2(S(\tau))>=\beta^2<S^2(\tau)>+<S(\tau)>\]
where
\[<S^2(\tau)>=\int_{0}^{\infty}x^2f_{S(\tau)}(x)dx.\]
For small $\tau$ the last characteristic behaves:
\[<S^2(\tau)>=\int_{0}^{\infty}x^2\frac{1}{\alpha x}\tau f_{T(x)}(\tau)dx=\int_{0}^{\infty}\frac{x}{\alpha x^{1/\alpha} }\tau e^{-\lambda \tau+\lambda^{\alpha}x}f_{U(1)}(\tau/x^{1/\alpha})dx=\int_{0}^{\infty}\frac{\tau^{2\alpha}}{x^{2\alpha}}e^{-\lambda \tau+\lambda^{\alpha}(\tau/x)^{\alpha}}f_{U(1)}(x)dx\sim \tau^{2\alpha}.\]
On the other hand for large $\tau$ we have:
\[<S^2(\tau)>=\tau^{2\alpha}e^{-\lambda\tau}\int_{0}^{\infty} x^{-2\alpha}e^{\lambda^{\alpha}(\tau/x)^{\alpha}}f_{U(1)}(x)dx.\]

Therefore as a final result we obtain that the ensemble averaged MSD for the process $\{Y_{S}(\tau)\}$ satisfies:
\[<Y^2_S(\tau)>\sim c_1\tau^{2\alpha}+c_2\tau^{\alpha},~~\mbox{when}~~\tau\rightarrow0\]\[<Y^2_S(\tau)>\sim d_1\tau^{2\alpha}e^{-\lambda\tau}\int_{0}^{\infty} x^{-2\alpha}e^{\lambda^{\alpha}(\tau/x)^{\alpha}}f_{U(1)}(x)dx+d_2\tau, ~~\mbox{when}~~\tau\rightarrow \infty\]
for some constants $c_1$, $c_2$, $d_1$, $d_2$. Let us mention $c_1$ and $d_1$ depend on $\beta$ in such way that for $\beta=0$, $c_1=d_1=0$, therefore for $\beta=0$ the ensemble averaged MSD behaves like $\tau^{\alpha}$ for small values of $\tau$ and like $\tau$ for large values of this parameter, \cite{stanis}.\\

The ensemble averaged MSD calculated on the basis of $1000$ trajectories for the process $\{Y_{S}(\tau),~\tau\geq 0\}$ we present on the bottom panel of Fig. \ref{msd}. To the simulation we take $\alpha=0.8$, $\lambda=1$ and $\beta=0.01$. We show also the fitted power function for small and large values of arguments. Let us mention that on the top panel we present the ensemble averaged MSD for the ABM with tempered stable subordinator $\{Y_T(t),~t\geq 0\}$ with fitted polynomial of order $2$.
\subsection{Estimation procedure}
\begin{figure}[tb]
\begin{center}
\includegraphics[width=9cm]{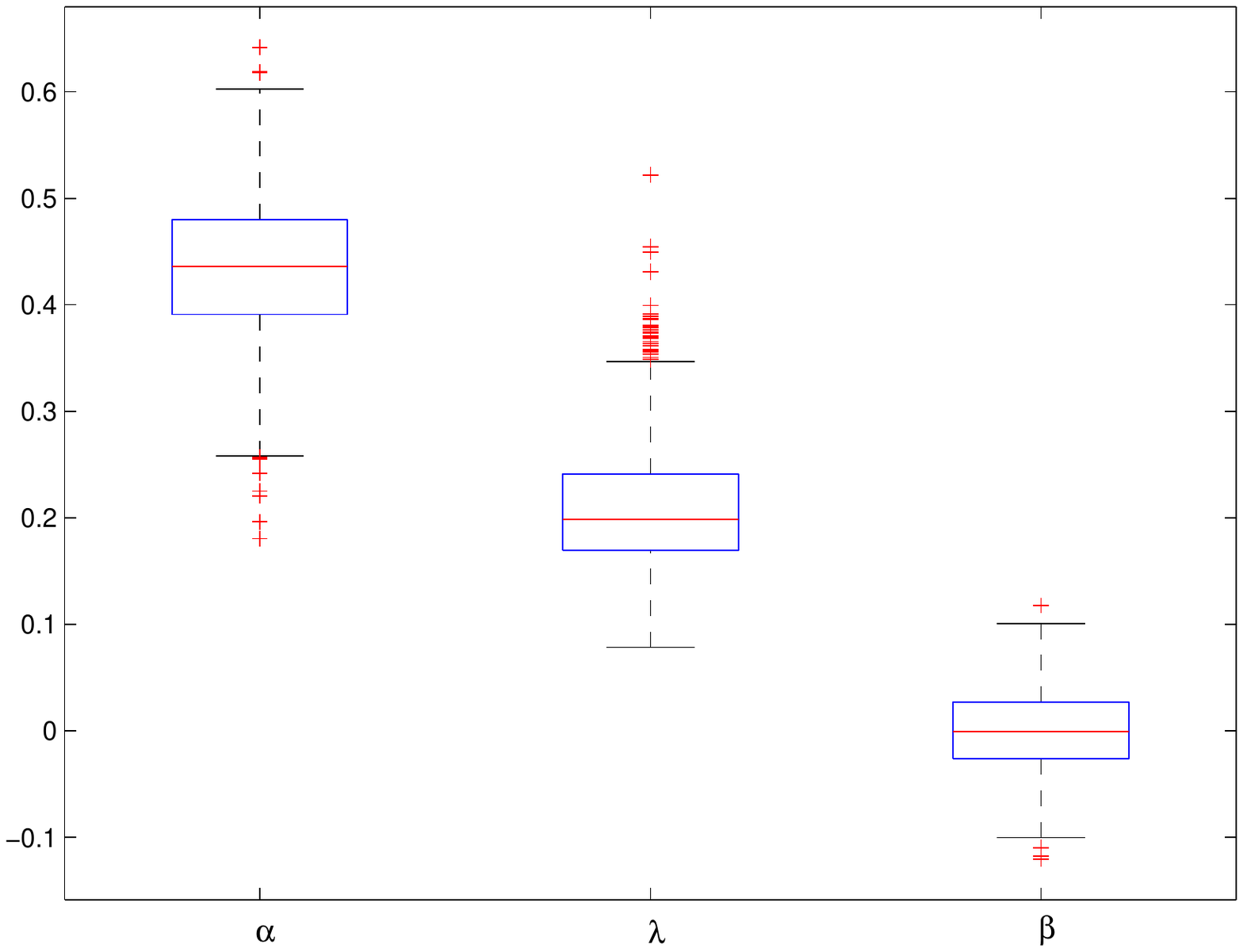} \caption{The boxplots of the values of estimators for parameters of ABM with inverse tempered stable subordinator $\{Y_S(\tau)\}$. The values were calculated on the basis of $1000$ trajectories of length $1000$ each. The theoretical values are : $\alpha=0.4$, $\lambda=0.2$, $\beta=0$.}\label{esti_2}
\end{center}
\end{figure}

The estimation procedure for parameters $\alpha$, $\beta $ and $\lambda$ of ABM with inverse tempered stable subordinator is partially described in \cite{orzelwyl}. We only mention here that it is based on the decomposition of the time series into two vectors. The first one is responsible for the lengths of constant time periods typical for the subdiffusive processes, while the second one appears after removing  the constant periods. According to the theory, in our case the first vector constitutes independent identically distributed sample from tempered stable distribution. Because the right tail of the tempered stable distribution behaves like $e^{-\lambda x}x^{-\alpha}$ therefore we propose to estimate the parameters $\alpha$ and $\lambda$ by fitting this function to the empirical tail by using the least squares method.  The second vector is related to the external process $\{X(t)\}$, therefore in the case of ABM the differenced series has Gaussian distribution with mean $\beta$. Therefore the parameter we estimate as a mean of the differenced series. \\In order to present the efficiency of the presented method  we simulate $1000$ trajectories (each of length $1000$) of  ABM with inverse tempered stable subordinator $\{Y_S(\tau)\}$ with $\alpha=0.4$, $\lambda=0.2$ and $\beta=0$. Next, we estimate the parameters by using the presented scheme and in Fig. \ref{esti_2} we show the results. As we observe the fitted parameters are close to the theoretical values.

\section{Applications}\label{appl}
\begin{figure}[tb]
\begin{center}
\includegraphics[width=9cm]{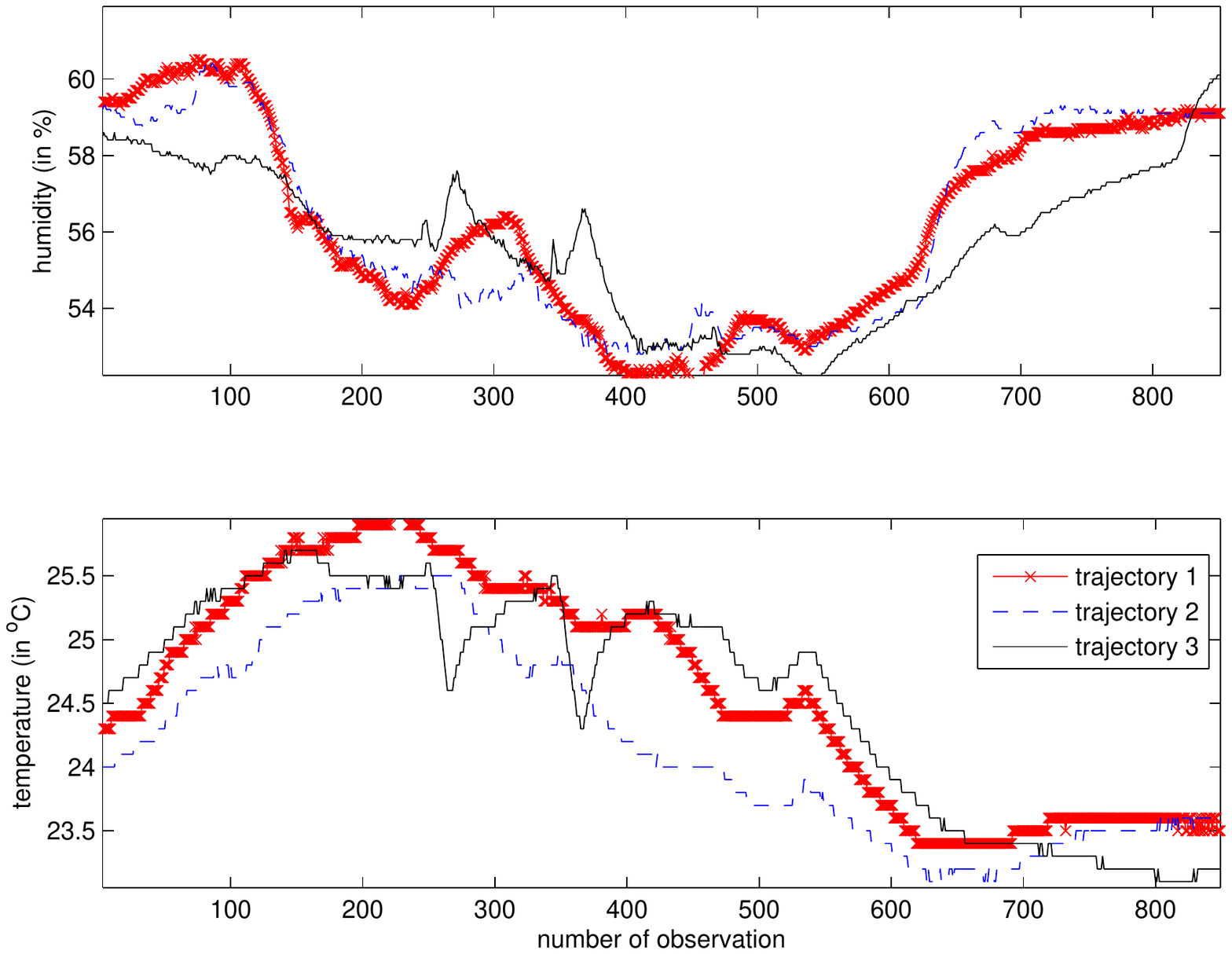} \caption{The humidity (top panel) and temperature (bottom panel) measured by three sensors between 9:48 and 23:59 on 12.08.2010.  }\label{real1}
\end{center}
\end{figure}

In this section we examine real data sets that describe humidity (in $\%$) and temperature (in $^oC$) of the indoor air in some open space of huge company. For simplicity we denote humidity as DATA1 and temperature - as DATA2. Those two analyzed quantities were measured by three sensors placed in the open space. Therefore for DATA1 and DATA2 we have three paths (trajectories). 
The humidity and temperature was quoted per minute. To the analysis we take data from 9:48--23:59 on 12.08.2010 ($851$ observations of each trajectory). On Fig. \ref{real1} we present paths of 
humidity (top panel) and  data related to temperature (bottom panel).

\begin{figure}[tb]
\begin{center}
\includegraphics[width=9cm]{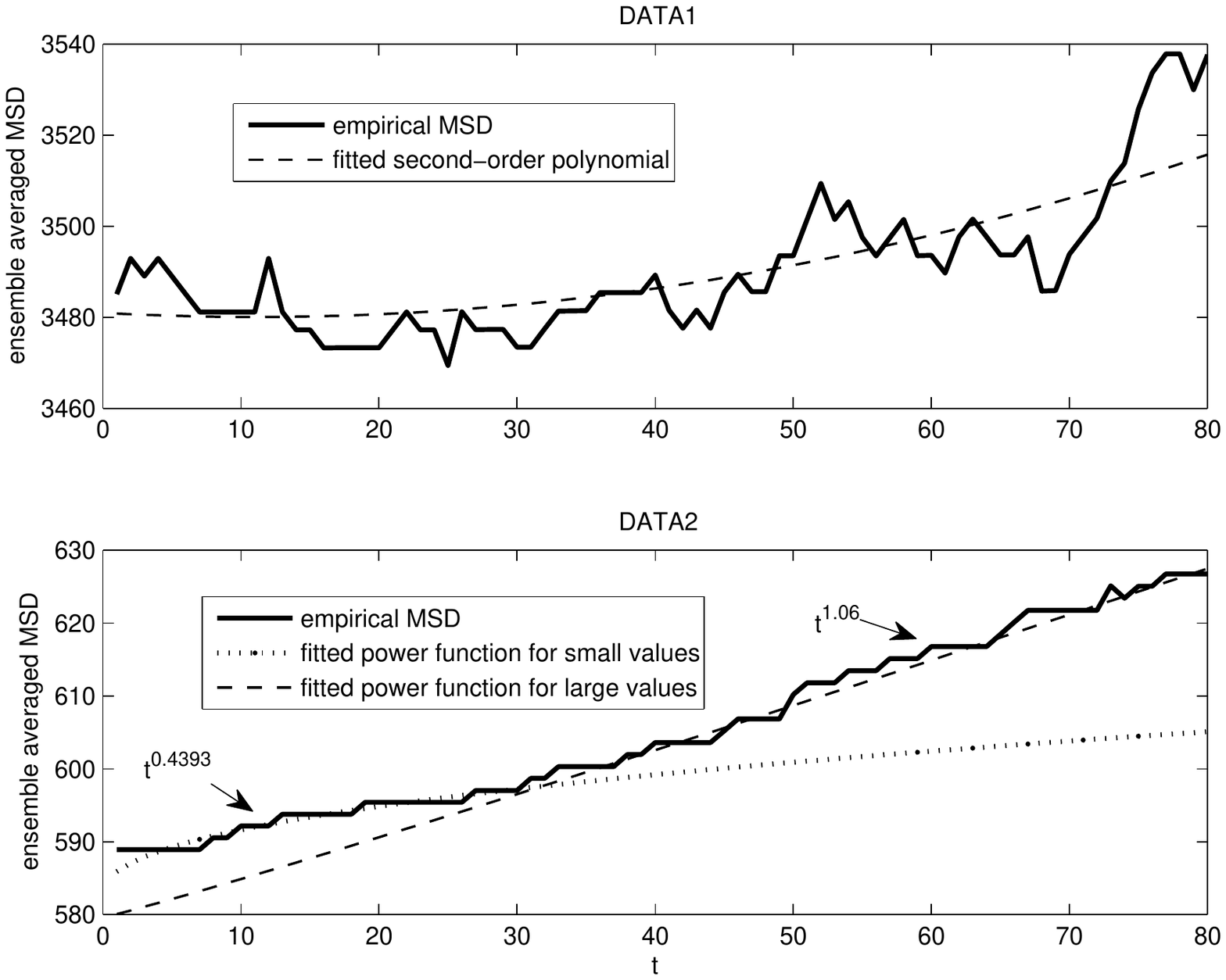} \caption{The ensemble averaged MSD of humidity (top panel) and temperature (bottom panel) together with the fitted second-order polynomial (for DATA1) and power function  for small and large arguments (for DATA2).}\label{msd_real}
\end{center}
\end{figure}

As we observe on Fig. \ref{real1}, DATA2 exhibit behavior typical to subdiffusive process $\{Y_S(\tau)\}$ described in previous section, namely the constant time periods. This behavior is not visible for DATA1 (top panel).   In the first step we examine the empirical ensemble averaged MSD calculated on the basis of three trajectories (of DATA1 and DATA2), see Fig. \ref{msd_real}. To the ensemble averaged MSD of DATA1 we fit the second-order polynomial (top panel), while for small and large values of the MSD of DATA2 - the power functions. In both cases we use the least squares method.
As we observe, the ensemble averaged MSD for DATA1 behaves like second-order polynomial  for all arguments that  suggests behavior similar to this observed in Fig. \ref{msd} (top panel). The empirical MSD of the temperature that we observe on bottom panel of Fig. \ref{msd_real} clearly suggests subdiffusive behavior. 

\begin{figure}[tb]
\begin{center}
\includegraphics[width=9cm]{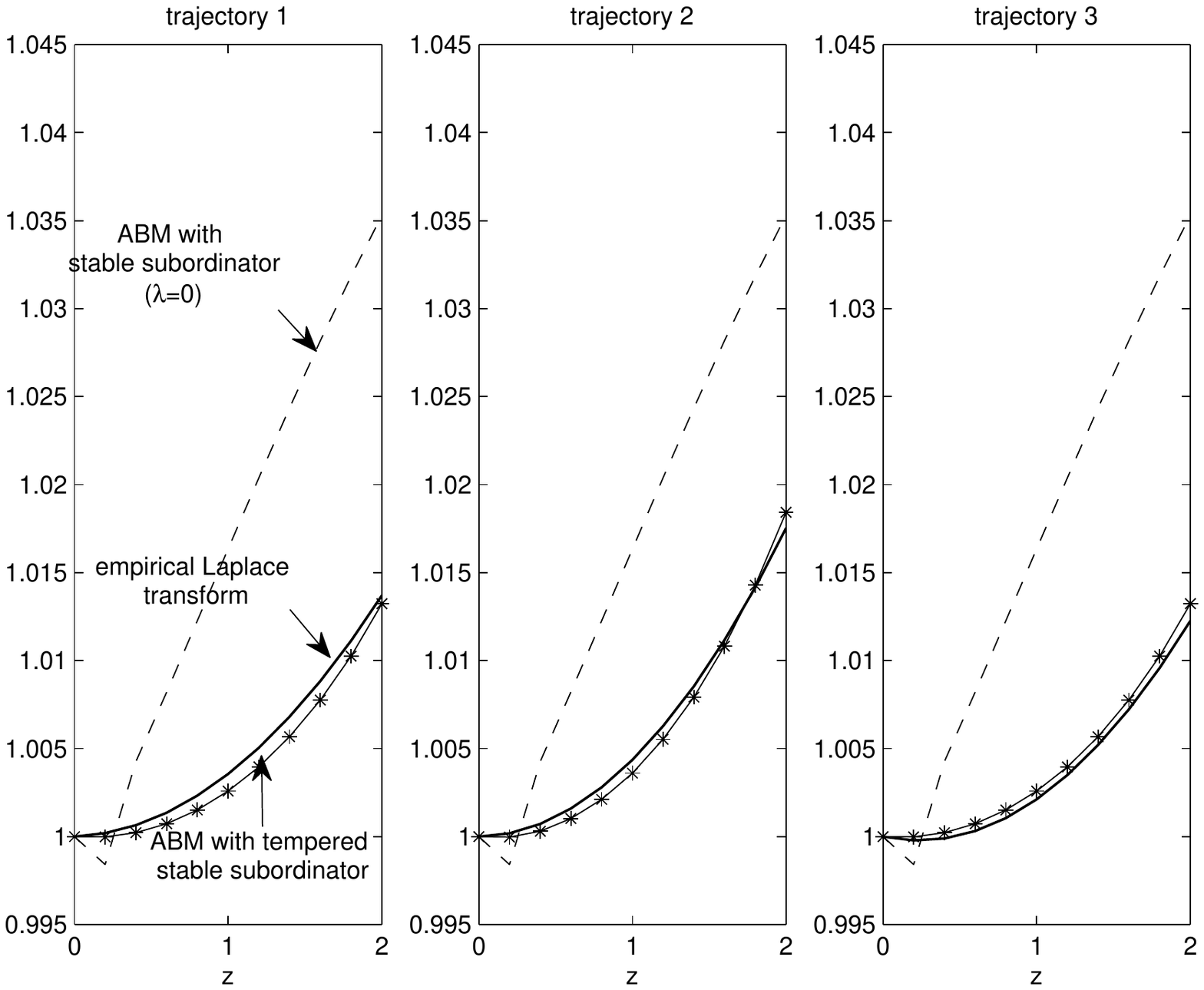} \caption{The empirical Laplace transform (solid line), theoretical Laplace transform for ABM with tempered stable subordinator (star line) and theoretical Laplace transform for ABM with stable subordinator (dashed line) for three trajectories of DATA1.}\label{LT_real}
\end{center}
\end{figure}

In the next step of our analysis we estimate the parameters of the suggested processes, i.e. ABM with tempered stable subordinator for DATA1 and ABM driven by inverse tempered stable subordinator for DATA2. In both cases we use the procedures presented in sections \ref{TSS} and \ref{ITSS}, respectively. In the Table \ref{tab1} we present the estimated parameters for DATA1.
\begin{table}[bh!]
\label{tab1}
\centering
\begin{tabular}{|c|c|c|c|}
\hline
 DATA1& $\hat{\alpha}$  &   $\hat{\lambda}$   &   $\hat{\beta}$ \\\hline
\hline
trajectory 1  & $0.23$ &$6.1$   &$0.11$      \\\hline

trajectory 2  &$0.29$ &$5.9$   &$0.12$      \\\hline 
trajectory 3  &$0.24$ & $6.1$  & $0.12$     \\\hline
  
\end{tabular}
\caption{Estimated parameters of the ABM with tempered stable subordinator for DATA1.}\label{tab1}
\end{table}\\
Moreover in Fig. \ref{LT_real} we present the empirical Laplace transform for each of three trajectories of DATA1 and theoretical one given in (\ref{ltYT}). In order to confirm that the tempered stable subordinator is better than the stable one we show also the Laplace transform for the process $\{Y_T(t)\}$ with $\lambda=0$ (the other parameters we estimate by using the same method as this presented in section \ref{TSS}). Moreover  we test also the Gaussian and $\alpha-$stable distributions for differenced series of DATA1 by using the methods based on the distance between empirical nad theoretical distribution functions,\cite{krzysiek}. Those tests reject the hypothesis of the stable or Gaussian behavior of the observed data.
\\
After analysis of the humidity we start the examination of temperature. The ensemble averaged MSD indicates the data can be considered as subdiffusive process. Therefore we propose to use the ABM with inverse tempered stable subordinator described in section \ref{ITSS}. Let us mention that the $\alpha-$stable subordinator is not appropriate in this case because of the behavior of ensemble averaged MSD that in stable case behaves like $\tau^{\alpha}$ for all values of arguments.    The estimated values of $\alpha$, $\lambda$ and $\beta$ parameters for three trajectories we present in Table \ref{rela_22}.

\begin{table}[bh!]
\label{rela_22}
\centering
\begin{tabular}{|c|c|c|c|}
\hline
 DATA2& $\hat{\alpha}$  &   $\hat{\lambda}$   &   $\hat{\beta}$ \\\hline
\hline
trajectory 1  & $0.41$ &$0.13$   &$-0.0085$      \\\hline

trajectory 2  &$0.3936$ &$0.11$   &$-0.0039$      \\\hline 
trajectory 3  &$0.4906$ & $0.129$  & $-0.012$     \\\hline
  
\end{tabular}
\caption{Estimated parameters of the ABM with inverse tempered stable subordinator for DATA2.}\label{rela_22}
\end{table}

\section{Conclusions}

In this paper we have examined two processes related to subordinated ABM driven by tempered stable type systems. The first one, so called normal tempered stable, arises after subordination of ABM by strictly increasing tempered stable process, while the second one is a result of subordination of ABM with inverse tempered stable model. We have compared the main characteristics of such  systems, like Laplace transforms, asymptotic behavior of pdf as well as the ensemble averaged MSD, that can be an useful tool of recognition between considered models. We have described also the estimation procedures for the parameters of considered processes and validate them.  Finally, we have analyzed the real data related to indoor air quality in context of presented methodology. 
\begin{center}
\section*{Acknowledgements}
\end{center}
 I would like to thank Laboratory of Sensor Techniques and Indoor Air Quality
Research from Wroc{\l}aw University of Technology for access to real data base. \\\\

\end{document}